**Reply to Note on cond-mat/0510270: Jarzynski equation for adiabatically stretched rotor**

Jaeyoung Sung

Department of Chemistry, Chung-Ang University, Seoul 156-756, Korea

**Abstract**

Although the analysis in cond-mat/0510270 is correct, this doesn't mean Jarzynski relation holds always for an arbitrary process. There exists a sufficient and necessary condition for Jarzynski relation to hold for an adiabatic parameter switching process. In contradiction to recent assertions, the validity condition of Jarzynski relation for an adiabatic process is not always satisfied.

In Ref. 1, Jarzynski proposed a relation that relates free energy difference between two macrostates of a system to statistical distribution of work done on the system during an arbitrary process connecting the two macrostates. Recently, it was asserted that Jarzynski relation always holds if the phase space extension of an initial state of the system is sampled completely according to Boltzmann distribution.[2] In accordance with the latter assertion Bier showed in Ref. 3 that Jarzynski relation holds for a particle confined on a sphere with a temporally varying radius $r$ whose dynamics is governed by Hamilton's equation of motion with Hamiltonian

$$H_r(\theta,\phi,p_\theta,p_\phi) = \frac{1}{2\mu r^2}\left[p_\theta^2 + \left(\frac{p_\phi}{\sin^2\theta}\right)^2\right] + U(r), \qquad (1)$$

where $\mu$ and $U$ respectively denote the mass of the particle and the potential dependent on $r$ only. In Eq. (1) $\theta$ and $\phi$ are the polar angle and the azimuthal angle of the position vector of the particle in the spherical polar coordinate system, and $p_\theta$ and $p_\phi$ are the associated momenta. The result of Ref. 3 is in contradiction with that of Ref. 4 in which it is mistakenly assumed that the kinetic energy of the particle given by the first term on the R.H.S. of Eq. (1) conserves during an adiabatic process changing $r$ from $r_0$ to $r_1$ by a central force and the work done $W$ on the system during the adiabatic process is equal to $W = U(r_1) - U(r_0)$.

However, in Ref. 5, a revised version of Ref. 4, the mistake was remedied by considering a slightly different system. The latter system is composed of a rigid rotor and a spring that connects the coordinate origin to the center of mass of the rigid rotor, an example of which system is depicted in Fig. 1. If the system assumes a configuration with the center of mass of the rigid rotor held fixed at $\mathbf{R}$, the Hamiltonian of the system in the configuration is given by

$$H(\theta,\phi,p_\theta,p_\varphi,\mathbf{R}) = \frac{1}{2I}\left[p_\theta^2 + \left(\frac{p_\phi}{\sin^2\theta}\right)^2\right] + U(|\mathbf{R}|), \tag{2}$$

where $I$ denotes the moment of inertia of the rigid rotor, constant in time, and $U$ denotes the potential energy due to the spring part of the system, dependent on the magnitude $|\mathbf{R}|$ of the position vector $\mathbf{R}$ of the center of mass of the rigid rotor. In Eq. (2), ($p_\theta$, $p_\phi$, $\theta$, $\phi$) designates the generalized momenta and coordinates describing a rotational motion of the rigid rotor. Note that the kinetic energy for the rotation of the rigid rotor given by the first term on the R.H.S. of Eq. (2) is independent of $\mathbf{R}$ so that it conserves during any adiabatic process that changes $\mathbf{R}$. For the latter model, free energy difference between the equilibrium configurational state of the system with $|\mathbf{R}|=R_0$ and that with $|\mathbf{R}|=R_1$ turns out different from the prediction of Jarzynski relation.[5] We can draw the similar conclusion for a number of other models, which cannot be covered completely here.

However, the latter example along with the expansion of ideal gas into vacuum show that Jarzynski relation does not hold always.[6,7] As a matter of fact, there exists a sufficient and necessary condition for Jarzynski relation to hold for a Hamiltonian system.[6] In the following, we will provide some detailed explanation about the validity condition of Jarzynski relation for an adiabatic process during which dynamics of a system obeys classical dynamics by considering a couple of different pairs of configurational states of the system considered in Ref. 5. In the analysis we assume that the entire phase space of an initial equilibrium state of the system can be sampled completely. We will finish this reply by confirming that the validity condition of Jarzynski relation is indeed satisfied for the model considered in Ref. 3, a particle confined on a sphere.

Let us begin our discussion by noting that the original derivation of Jarzynski relation for an adiabatic process assumes neither a specific form of system Hamiltonian nor a specific type of the state parameter defining a macrostate of the system.[1] Instead, the derivation in Ref. 1 of Jarzynski relation for an adiabatic process invokes two requirements: 1) before the adiabatic process, probability distribution of a microstate of a system is given by the Boltzmann distribution with an arbitrary state parameter; and 2) a system evolves according to classical dynamics during the adiabatic process. However, even when the latter requirements are satisfied, Jarzynski relation for an adiabatic process may not hold unless the following validity condition of Jarzynski relation is satisfied: *Jarzynski relation holds for an adiabatic process if and only if the phase space extension of a system state prepared at the very end of the adiabatic process, in which value of a state parameter of the system is changed in a time interval, coincides with the phase space extension of the thermal equilibrium state of the system with the state parameter having the final value.*[6]

To expose the concept of the latter statement concretely, we will examine the free energy difference, the validity condition of Jarzynski relation, and the prediction of Jarzynski relation, for a couple of different pairs of macrostates of the system considered in Ref. 5, i.e. freely rotating rigid rotor and a spring that connects our coordinate origin to the center of mass of the rigid rotor, depicted in Fig. 1. The Hamiltonian of the system without any constraint on dynamical variables is given by

$$H = \frac{1}{2I}(p_\theta^2 + \frac{1}{\sin^2\theta} p_\phi^2) + \frac{|\mathbf{P}|^2}{2M} + U(|\mathbf{R}|), \tag{4}$$

where $\mathbf{P}$ and $M$ denote the momentum conjugated to $\mathbf{R}$ and the total mass of the rigid rotor, respectively. Other notations in Eq. (4) are the same as those in Eq. (2).

A macrostate of a system is composed of an ensemble of microstates of the system consistent with a set of constraints that we choose to identify the macrostate. As is well known, a constraint required to construct a canonical ensemble is conservation of total energy $E$ of the entire ensemble, and among various possible distribution of a microstate of the system in the ensemble consistent with the constraint on the total energy, the Boltzmann distribution is the most probable distribution or the equilibrium distribution, given by

$$f_{eq}(\Gamma) = \exp[-\beta H(\Gamma)]/q(\beta), \quad (5)$$

with $\beta = 1/k_B T$ and $q(\beta) = \int d\Gamma \exp(-\beta H(\Gamma))$. Here $\Gamma$ denotes a vector specifying a microstate of our system, i.e. $\Gamma = (p_\theta, p_\phi, \theta, \phi, \mathbf{P}, \mathbf{R})$. If $U$ in Eq. (4) is dependent on external parameters as well, so are the molecular partition function $q$ and the free energy defined by $A(\beta) = -k_B T \ln q(\beta)$. An example of such external parameter is volume of the container, if any, containing our system. If our system were composed of a number $N$ of identical subsystems, $N$ would be also one of the external parameters. An equilibrium state of the system is identified by values of such external parameters and the absolute temperature, $T$, originating from the constraint for the total energy of the canonical ensemble.

However, there are situations where we are interested in a macrostate of a system with an additional set of constraints on dynamical variables of the system. For example, it is often of interest to a biophysicist to estimate equilibrium free energy difference between configurational states of a biopolymer. To represent a configurational state of a polymer chain, one may use, for instance, the end-to-end distance vector, $\mathbf{R}_{ETE}$, of the polymer chain.[8] The equilibrium configurational state of a polymer chain with $\mathbf{R}_{ETE}$ being a constant vector, say **C,** at temperature $T$ is composed of canonical ensemble of microstates of the polymer chain whose end-to-end vector $\mathbf{R}_{ETE}$ is given by C, constant in time, $d\mathbf{R}_{ETE}/dt = 0$. A variety of different configurational states of the polymer chain can be defined by choosing different constraints on dynamical variables of the polymer chain.[8] For example, one can identify a configurational state of the polymer chain by the magnitude $|\mathbf{R}_{ETE}|$ of $\mathbf{R}_{ETE}$ or the radius of gyration, $R_G$ of the polymer. Each one of $\mathbf{R}_{ETE}$, $|\mathbf{R}_{ETE}|$, and $R_G$ can be a state parameter for identifying a configurational state of the polymer chain. A phase-space extension and a free energy of the polymer chain in a configurational state are dependent on a type and value of the state parameter that we choose to identify the configurational state of the polymer chain.

For the system with Hamiltonian given by Eq. (4), among a variety of possibilities, let us consider the equilibrium configurational state $(\beta, \mathbf{P}=0, \mathbf{R}=\mathbf{r})$ of the system defined by the following constraints, $\mathbf{P}=0$ and $\mathbf{R}=\mathbf{r}$ with $\mathbf{r}$ being a constant vector, in addition to the constraint for the total energy of the ensemble. The equilibrium distribution of the system consistent with these constraints is given by

$$f'_{eq}(\Gamma') = \frac{\exp[-\beta h_{\mathbf{r}}(\Gamma')]}{q(\beta, \mathbf{r})}, \tag{6}$$

where $\Gamma'$ is the vector specifying a microstate of the rigid rotor system with the constraints, $\mathbf{P}=0$ and $\mathbf{R}=\mathbf{r}$, i.e. $\Gamma' = (p_\theta, p_\phi, \theta, \phi, \mathbf{P}=0, \mathbf{R}=\mathbf{r})$. In Eq. (6), $h_{\mathbf{r}}$ and $q$ are given by

$$h_{\mathbf{r}}(\Gamma') = \frac{1}{2I}(p_\theta^2 + \frac{1}{\sin^2\theta}p_\phi^2) + U(|\mathbf{r}|) \tag{7}$$

and

$$q(\beta, \mathbf{r}) = \int d\Gamma' \exp[-\beta h_{\mathbf{r}}(\Gamma')], \tag{8}$$

respectively. In Eq. (8) and after, $\int d\Gamma'$ stands for $\int_{-\infty}^{\infty} dp_\theta \int_{-\infty}^{\infty} dp_\phi \int_0^\pi d\theta \int_0^{2\pi} d\phi$. $q(\beta,\mathbf{r})$ denotes the molecular partition function of the system in state $(\beta, \mathbf{P}=0, \mathbf{R}=\mathbf{r})$. Therefore, equilibrium free energy difference $\Delta A$ between the system in state $(\beta, \mathbf{P}=0, \mathbf{R}=\mathbf{r}_1)$ and that in state $(\beta, \mathbf{P}=0, \mathbf{R}=\mathbf{r}_2)$ is given by

$$\begin{aligned}\Delta A &\equiv A(\beta, \mathbf{R}=\mathbf{r}_2, \mathbf{P}=0) - A(\beta, \mathbf{R}=\mathbf{r}_1, \mathbf{P}=0)\\ &= -\beta^{-1} \ln[q(\beta,\mathbf{r}_2)/q(\beta,\mathbf{r}_1)] \\ &= U(|\mathbf{r}_2|) - U(|\mathbf{r}_1|)\end{aligned} \qquad (9)$$

For this model the above-mentioned validity condition of Jarzynski relation is satisfied. In order to estimate free energy difference between state $(\beta, \mathbf{P}=0, \mathbf{R}=\mathbf{r}_1)$ and state $(\beta, \mathbf{P}=0, \mathbf{R}=\mathbf{r}_2)$ from Jarzynski relation, let us consider the following process connecting the two states. Let the system be initially prepared in state $(\beta, \mathbf{P}=0, \mathbf{R}=\mathbf{r}_1)$ in thermal equilibrium with a heat bath. At time zero, we isolate the system from the heat bath, and begin an adiabatic process to change the value of $\mathbf{R}$ from $\mathbf{r}_1$ to $\mathbf{r}_2$ by an external force in time interval $[0, t_S]$. After the completion of the adiabatic process we get the system in contact with the heat bath again and let the system relax to the final equilibrium state $(\beta, \mathbf{P}=0, \mathbf{R}=\mathbf{r}_2)$. Later we will turn to the fact that the statistical distribution of work done on the system during the overall process is the same as that during the adiabatic process, as the latter thermal relaxation process does not cost any work. Here, the phase space extension of the system in equilibrium configurational state $(\beta, \mathbf{P}=0, \mathbf{R}=\mathbf{r})$ is given by $\mathbf{P}=0$ and $\mathbf{R}=\mathbf{r}$, for the center of mass degrees of freedom and by $-\infty < p_\theta, p_\phi < \infty$, $0 \leq \theta \leq \pi$, and $0 \leq \phi < 2\pi$ for the degrees of freedom for the internal rotation of the rigid rotor. Note that the phase-space extension for $(p_\theta, p_\phi, \theta, \phi)$ of the system in state

$(\beta, \mathbf{P} = 0, \mathbf{R} = \mathbf{r})$ are the same at any value of $\mathbf{r}$; in other words, it does not change with $\mathbf{r}$. Now let us consider the phase space extension of the system at the very end of the adiabatic process mentioned above. For our model, the Hamilton's equation of motion governing dynamics of $(p_\theta, p_\phi, \theta, \phi)$ is independent of $\mathbf{r}$, so the initial phase-space extension for $(p_\theta, p_\phi, \theta, \phi)$ is not perturbed by the adiabatic process. In other words, throughout the adiabatic process in which $\mathbf{R}$ is changed from $\mathbf{r}_1$ to $\mathbf{r}_2$, the phase space extension for $(p_\theta, p_\phi, \theta, \phi)$ is the same with that of the initial thermodynamic state $(\beta, \mathbf{P} = 0, \mathbf{R} = \mathbf{r}_1)$. Remembering that the phase space extension for $(p_\theta, p_\phi, \theta, \phi)$ of the system in equilibrium state $(\beta, \mathbf{P} = 0, \mathbf{R} = \mathbf{r})$ is independent of $\mathbf{r}$, one can see that the phase space extension for $(p_\theta, p_\phi, \theta, \phi)$ of the system state with $\mathbf{P}=0$ and $\mathbf{R} = \mathbf{r}_2$ at the very end of the adiabatic process coincides with that of the system in configurational state $(\beta, \mathbf{P} = 0 \mathbf{R} = \mathbf{r}_2)$; therefore, the above-mentioned validity condition of Jarzynski relation is satisfied.

Let us confirm that Jarzynski relation provides the correct equilibrium free energy difference between two configurational states, $(\beta, \mathbf{P} = 0, \mathbf{R} = \mathbf{r}_1)$ and $(\beta, \mathbf{P} = 0, \mathbf{R} = \mathbf{r}_2)$ by direct comparison between the exact result given by Eq. (9) and the predicted result of Jarzynski relation. According to Ref. 1, the estimation $\Delta A_J$ of Jarzynski relation for free energy difference between two states should be related to the distribution of work $W$ done on our system during an arbitrary process connecting the two states by

$$\exp(-\beta \Delta A_J) = \langle \exp(-\beta W) \rangle. \tag{10}$$

For the adiabatic process we consider here, the R.H.S. of Eq. (10) is given by

$\langle \exp(-\beta W) \rangle = \int d\Gamma' \exp[-\beta W(\Gamma')] f'_{eq}(\Gamma')$ where $f'_{eq}$ denotes the equilibrium distribution of a microscopic state $\Gamma'$ of the system in state $(\beta, \mathbf{P} = 0, \mathbf{R} = \mathbf{r}_1)$, and $W(\Gamma')$ denotes the work done on the system with initial state at $\Gamma'$ during the adiabatic process. To evaluate $\Delta A_J$, one should repeat the above-mentioned adiabatic process for every initial microstate $\Gamma'$ of the system in the initial state $(\beta, \mathbf{P} = 0, \mathbf{R} = \mathbf{r}_1)$. The amount of work $W(\Gamma')$ done on the system during the adiabatic process changing $\mathbf{R}$ from $\mathbf{r}_1$ to $\mathbf{r}_2$ in a time interval is equal to $W(\Gamma') = U(\mathbf{r}_2) - U(\mathbf{r}_1)$ for any initial microstate $\Gamma'$ in the initial state $(\beta, \mathbf{P} = 0, \mathbf{R} = \mathbf{r}_1)$, because the rotational energy of the rigid rotor given by the first term in the R.H.S. of Eq. (7) conserves during the adiabatic process according to classical dynamics. Therefore, we obtain $\langle \exp(-\beta W) \rangle = \exp(-\beta \Delta A)$ with $\Delta A$ being the correct result given by Eq. (9) to confirm that $\Delta A_J = \Delta A$ or Jarzynski relation holds for this case where the validity condition of Jarzynski relation is satisfied.

However, the validity condition of Jarzynski relation is not always satisfied. For example, let us consider a configurational state $(\beta, \mathbf{P} = 0, |\mathbf{R}| = R)$ of the system defined by the following constraints, $\mathbf{P} = 0$ and $|\mathbf{R}| = R$ with $R$ being a constant, in addition to the constraint for the total energy of the ensemble. Then the equilibrium distribution of a microstate of the system with the latter constraints is given by

$$f''_{eq}(\Gamma'') = \frac{\exp[-\beta h'_r(\Gamma'')]}{q'(\beta, r)}, \qquad (11)$$

where $\Gamma''$ is a vector specifying a microstate of the system consistent with the just-mentioned constraints, i.e., $\Gamma'' = (p_\theta, p_\phi, \theta, \phi, \mathbf{P} = 0, \mathbf{R})$ with $|\mathbf{R}| = R$. In Eq. (11), $h'_r$ and $q'$ are respectively given by

$$h'_r(\Gamma'') = \frac{1}{2I}(p_\theta^2 + \frac{1}{\sin^2\theta}p_\phi^2) + U(|\mathbf{R}|), \tag{12}$$

and

$$\begin{aligned}q'(\beta,r) &= \int_R d\Gamma'' \exp[-\beta h'_r(\Gamma'')] \\ &\equiv S_d R^{d-1} \exp[-\beta U(R)] \int d\Gamma' \exp\left(-\beta \frac{L^2}{2I}\right),\end{aligned} \tag{13}$$

where $\int_R d\Gamma''$ stands for $\int_{-\infty}^{\infty} dp_\theta \int_{-\infty}^{\infty} dp_\phi \int_0^\pi d\theta \int_0^{2\pi} d\phi \int d\mathbf{P} \int d\mathbf{R} \, \delta(\mathbf{P})\delta(|\mathbf{R}|-R)$. $S_d$ denotes the surface area of the $d$-dimensional sphere with a unit radius, i.e. $S_d = \frac{2\pi^{d/2}}{\Gamma(d/2)}$, and $L^2/2I$ denotes the first term on the R.H.S. of Eq. (12) with $L$ being a magnitude of angular momentum of the rigid rotor, i.e. $L^2 = p_\theta^2 + \frac{1}{\sin^2\theta}p_\phi^2$. The free energy difference $\Delta A$ between the system in state $(\beta, \mathbf{P}=0, |\mathbf{R}|=R_1)$ and that in state $(\beta, \mathbf{P}=0, |\mathbf{R}|=R_2)$ is given by

$$\begin{aligned}\Delta A &= -\beta^{-1} \ln[q'(\beta,r_2)/q'(\beta,r_1)] \\ &= U(r_2) - U(r_1) - \beta^{-1}(d-1)\ln(r_2/r_2)\end{aligned}. \tag{14}$$

It is turn to examine whether or not the validity condition of Jarzynski relation is satisfied in this case. To estimate equilibrium free energy difference between state $(\beta, \mathbf{P}=0, |\mathbf{R}|=R_1)$ and state $(\beta, \mathbf{P}=0, |\mathbf{R}|=R_2)$ from Jarzynski relation, we consider the following process connecting the two states. Let the system be initially prepared in state $(\beta, \mathbf{P}=0, |\mathbf{R}|=R_1)$ in thermal equilibrium with a heat bath. At time zero, we isolate the system from the heat bath, and begin an adiabatic process to change the value of $|\mathbf{R}|$ from $R_1$ to $R_2$ by a central force in time interval $[0, t_S]$. After the completion of the adiabatic process we get the system in contact with the heat bath again and let the system relax to the final equilibrium state $(\beta, \mathbf{P}=0, |\mathbf{R}|=R_2)$. The phase space extension of the system in thermal equilibrium state $(\beta, \mathbf{P}=0, |\mathbf{R}|=R)$ is

given by $-\infty < p_\theta, p_\phi < \infty$, $0 \leq \theta \leq \pi$, $0 \leq \phi < 2\pi$ for the rotational degrees of freedom of the rigid rotor and by $P_j = 0$, $-r < R_j < r$ with constraint $\sqrt{\sum_{j=1}^{d} R_j^2} = R$ for the center of mass degrees of freedom of the rigid rotor. Here $P_j$ and $R_j$ respectively designate the j-th component of $\mathbf{P}$ and $\mathbf{R}$ in the Cartesian coordinate system. Note that the equilibrium phase space extension for $(p_\theta, p_\phi, \theta, \phi)$ and $P_j$ does not change with R, but that for $R_j$ does. Now let us consider the phase space extension of the system at the very end of the adiabatic process. Throughout the adiabatic process in which the value of R increases from $R_1$ to $R_2$ in the radial direction in time interval $[0, t_S]$ by a central force exerted on the center of mass of the rigid rotor, the phase space extension of $(p_\theta, p_\phi, \theta, \phi)$ does not change for the same reason as before, which is given by $-\infty < p_\theta, p_\phi < \infty$, $0 \leq \theta \leq \pi$, and $0 \leq \phi < 2\pi$. The latter phase space extension coincides with the phase space extension of $(p_\theta, p_\phi, \theta, \phi)$ in the final state $(\beta, \mathbf{P} = 0, |\mathbf{R}| = r_2)$ of the system in thermal equilibrium. In comparison, the phase space extension of $(\mathbf{P}, \mathbf{R})$ at the end of the adiabatic process cannot coincide with that in state $(\beta, \mathbf{P} = 0, |\mathbf{R}| = r_2)$. This follows from Gibbs's principle of conservation in phase, which states that the volume of phase-space extension of an initial state of a system conserves as long as the system evolves according to classical dynamics. So, throughout the adiabatic process, the volume of phase space extension for the center of mass degrees of freedom is the same as that of the initial state $(\beta, |\mathbf{R}| = R_1, \mathbf{P} = 0)$, which is given by

$\int d\mathbf{P} \int d\mathbf{R} \delta(\mathbf{P}) \delta(|\mathbf{R}| - R_1) = S_d R_1^{d-1}$. On the other hand, the volume of phase space extension for the center of mass degrees of freedom in the final equilibrium state $(\beta, |\mathbf{R}| = R_2, \mathbf{P} = 0)$ is given by $\int d\mathbf{P} \int d\mathbf{R} \delta(\mathbf{P}) \delta(|\mathbf{R}| - R_2) = S_d R_2^{d-1}$. As the phase space extension for the center of mass degrees of freedom of the system at the very end of adiabatic process has a different volume from that of the system in the final equilibrium state $(\beta, |\mathbf{R}| = R_2, \mathbf{P} = 0)$, the former cannot coincide with the latter; that is to say, the validity condition of Jarzynski relation is not satisfied in this case. Accordingly, as shown in Ref. 5, the predicted result $\Delta A_J$ of Jarzynski relation for free energy difference between two configurational states, $(\beta, |\mathbf{R}| = R_1, \mathbf{P} = 0)$ and $(\beta, |\mathbf{R}| = R_2, \mathbf{P} = 0)$ is not in agreement with the exact result given by Eq. (14).

Note that kinetic energy for the center of mass of the rigid rotor is zero for every microstate in the ensemble constituting the initial configurational state $(\beta, |\mathbf{R}| = R_1, \mathbf{P} = 0)$, and, during the adiabatic process in which a central force is exerted on the center of mass of the rigid rotor to change $R$ from $R_1$ to $R_2$, the direction of motion of the center of mass occurs only in the radial direction according to classical dynamics.

In comparison, if our system is initially in the macrostate $(\beta, \mathbf{R} \cdot \mathbf{P} = 0, |\mathbf{R}| = R)$ defined by the following constraints, $\mathbf{R} \cdot \mathbf{P} = 0$ and $|\mathbf{R}| = R$ with $R$ being a constant, in addition to the usual constraint for the total energy of the ensemble, the motion of the center of mass of the rigid rotor occurs, in general, not only in the radial direction but also in the tangential direction during the adiabatic process in which a value of $R$ is changed, say, from $R_1$ to $R_2$. For the latter adiabatic process in the course of the path connecting state $(\beta, \mathbf{R} \cdot \mathbf{P} = 0, |\mathbf{R}| = r_1)$ to state $(\beta, \mathbf{R} \cdot \mathbf{P} = 0, |\mathbf{R}| = r_2)$, it can be shown

that the validity condition of Jarzynski relation is satisfied about which we will not present a more detailed analysis here.

Instead let us discuss the model considered in Refs. 3 and 4, i.e. a particle confined on a sphere with temporally varying radius $r$ whose Hamiltonian is given by Eq. (1). Bier showed in Ref. 3 that Jarzynski relation holds for the latter model. We will show that the validity condition of Jarzynski relation is indeed satisfied for the latter model. If $(r,T)$ designates a thermal equilibrium state of the particle on a sphere with radius $r$ at temperature $T$, the phase space extension of the particle in state $(r,T)$ is given by $-\infty < p_\theta, p_\phi < \infty$, $0 \leq \theta \leq \pi$, and $0 \leq \phi < 2\pi$ for any value of $r$, where the notation is the same as those in Eq. (1). The phase space extension conserves during the adiabatic process that changes the value of $r$, say, from $r_1$ to $r_2$ in time interval $[0, t_S]$ due to the following operator identity:

$$\int_{-\infty}^{\infty} dp_\theta(0) \int_{-\infty}^{\infty} dp_\phi(0) \int_0^\pi d\theta(0) \int_0^{2\pi} d\phi(0) = \int_{-\infty}^{\infty} dp_\theta(t_S) \int_{-\infty}^{\infty} dp_\phi(t_S) \int_0^\pi d\theta(t_S) \int_0^{2\pi} d\phi(t_S)$$

or the following identity $\left\| \dfrac{\partial [p_\theta(t_S), p(t_S), \theta(t_S), \phi(t_S)]}{\partial [p_\theta(0), p(0), \theta(0), \phi(0)]} \right\| = 1$, which results from Liouville theorem and holds as long as dynamics of the system obeys Hamilton's equation of motion.[9] Remembering that the phase space extension of the system in state $(r_1, T)$ is the same as that of the system in state $(r_2, T)$, one can see that the phase space extension of our system at the end of the adiabatic process or at time $t_S$ coincides with that of the system in state $(r_2, T)$ so that the validity condition of Jarzynski relation is satisfied and Jarzynski relation holds as shown in Ref. 3.

Although Biers analysis in Ref. 3 is correct, this doesn't mean Jarzynski relation holds always for an arbitrary process.[5,6] There exists a sufficient and necessary condition for Jarzynski relation to hold for an adiabatic parameter switching process throughout which dynamics of system obeys classical dynamics.[6] In contradiction to recent assertion,[2] the validity condition of Jarzynski relation for an adiabatic process may not be satisfied always even if entire phase space extension could be sampled completely. Up to now we investigated the validity of Jarzynski relation only for an adiabatic parameter switching process. However, it can be shown that Jarzynski relation does not hold always for other nonequilibrium process than an adiabatic process, during which heat can flow in or out of our system. The analysis supporting the latter assertion will be reported elsewhere.

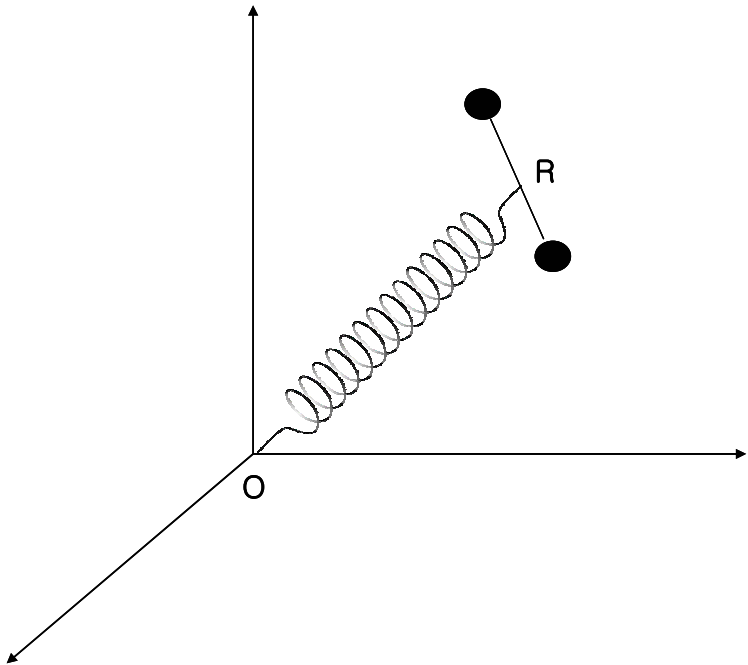

**Figure 1: The system of a rigid rotor and a spring that connects the coordinate origin, O, to the center of mass, R, of the rigid rotor considered in Ref. 5.**